\documentclass[aps,twocolumn,showpacs,prl,amsmath,amssymb,floatfix,superscriptaddress]{revtex4-1}

\usepackage{color}
\usepackage{bbm}
\usepackage{graphicx}
\usepackage{dcolumn}
\usepackage{bm}
\usepackage{array}
\usepackage{float}
\usepackage{supertabular}
\usepackage{longtable}
\usepackage{mathrsfs}
\usepackage{txfonts}
\usepackage{wasysym}
\usepackage[T1]{fontenc}
\usepackage[latin1]{inputenc}
\usepackage{amssymb}
\usepackage[usenames,dvipsnames]{xcolor}
\usepackage{amsmath}
\usepackage{amstext}
\usepackage{latexsym}
\usepackage[colorlinks=true,citecolor=Cerulean,linkcolor=RubineRed,urlcolor=Cerulean]{hyperref}

\begin{document}
\title{
 Anti-Kibble-Zurek Behavior in Crossing the Quantum Critical Point of a Thermally Isolated System Driven by a Noisy Control Field
}

\author{Anirban Dutta}
\affiliation{Department of Physics, University of Massachusetts, Boston, MA 02125, USA}
\author{Armin Rahmani}
\affiliation{Department of Physics and Astronomy and Quantum Matter Institute, University of British Columbia, Vancouver, British Columbia,  Canada V6T 1Z4}
\author{Adolfo del Campo}
\affiliation{Department of Physics, University of Massachusetts, Boston, MA 02125, USA}

\newcommand{\be}{\begin{equation}}
\newcommand{\ee}{\end{equation}}
\newcommand{\bea}{\begin{eqnarray}}
\newcommand{\eea}{\end{eqnarray}}

\def\q{{\bf q}}

\def\G{\Gamma}
\def\L{\Lambda}
\def\la{\lambda}
\def\g{\gamma}
\def\al{\alpha}
\def\s{\sigma}
\def\e{\epsilon}
\def\k{\kappa}
\def\ve{\varepsilon}
\def\l{\left}
\def\r{\right}
\def\te{\mbox{e}}
\def\d{{\rm d}}
\def\t{{\rm t}}
\def\K{{\rm K}}
\def\N{{\rm N}}
\def\H{{\rm H}}
\def\la{\langle}
\def\ra{\rangle}
\def\om{\omega}
\def\Om{\Omega}
\def\vep{\varepsilon}
\def\wh{\widehat}
\def\tr{{\rm Tr}}
\def\da{\dagger}
\def\iz{\left}
\def\zi{\right}
\newcommand{\beq}{\begin{equation}}
\newcommand{\eeq}{\end{equation}}
\newcommand{\beqa}{\begin{eqnarray}}
\newcommand{\eeqa}{\end{eqnarray}}
\newcommand{\intf}{\int_{-\infty}^\infty}
\newcommand{\into}{\int_0^\infty}

\begin{abstract}

We show that a thermally isolated system driven across a quantum phase transition by a noisy control field exhibits anti-Kibble-Zurek behavior, whereby slower driving results in higher excitations.
We characterize the density of excitations as a function of the ramping rate and the noise strength. The optimal driving time to minimize excitations is shown to scale as a universal power law of the noise strength. Our findings reveal the limitations of adiabatic protocols such as quantum annealing and demonstrate the universality of the optimal ramping rate.

\end{abstract}

\maketitle

Understanding adiabatic dynamics and its breakdown in many-body systems is fundamental to the progress of quantum technologies \cite{CZ12}. Adiabatic evolution is the cornerstone of the quantum annealing scheme for state preparation and quantum computation~\cite{Farhi,Suzuki}. The adiabatic theorem states that the dynamics of a physical system is free from diabatic transitions under slow driving  \cite{Born}. The suppression of excitations becomes  challenging in the absence of an energy gap, e.g., when crossing a quantum critical point (QCP) \cite{Dziarmaga10,Polkovnikov11,DZ13}. The density of excitation follows a universal power aw  as a function of the rate of change of the control field driving the system through the QCP \cite{Damski05,ZDZ05,Dziarmaga05,Polkovnikov05} and can be reduced by resorting to slow ramps. 
This universal scaling is the key prediction of the Kibble-Zurek mechanism (KZM), initially developed for classical continuous phase transitions \cite{Kibble76,Zurek96}.

While its experimental verification still calls for further studies \cite{DZ13}, KZM is believed to be broadly applicable. Yet, a conflicting observation has been reported in the study of mutiferroic systems: approaching the adiabatic limit, slower ramps generate more excitations \cite{Griffin12}. This counterintuitive phenomenon was termed anti-Kibble-Zurek (anti-KZ) dynamics. While tests of KZM in the quantum regime are scarce, the data in one of them hint at a possible anti-KZ behavior \cite{DeMarco11}. Here we show that in a  thermally isolated quantum system, the presence of noisy fluctuations in the control field naturally provides an explanation for  anti-KZ behavior.

We start by considering a linear passage through the QCP. A control field  $g$ is turned on from zero value to unity as in standard quantum annealing schemes, crossing a QCP at $g_c={1\over 2}$.  When the transition is crossed at a rate $1/\tau$ fixed by the ramp duration $\tau$,  KZM predicts  universal power-law for the density of excitations $n_0\propto \tau^{-\beta}$, with $\beta= d\nu/(1+z\nu)$, where $\nu$ and $z$ are the correlation length and dynamic critical exponents, and $d$ is the dimensionality of the system. The subindex in $n_0$ is introduced to denote noise-free driving. The density of excitations monotonically decreases with $\tau$ and vanishes in the limit of $\tau\to\infty$.
\begin{figure}[t]
\centerline{\includegraphics[width=80mm]{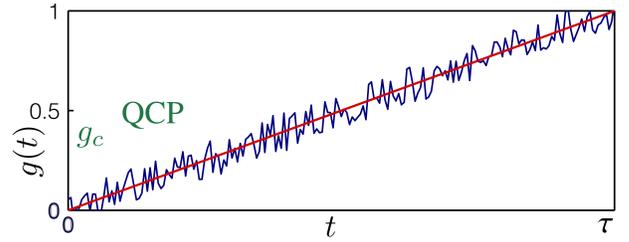}}
\caption{{\bf Schematic  driving through a QCP with a noisy control field.} Under an idealized smooth  control field $g(t)$, diabatic transitions are suppressed as the ramp time $\tau$ is increased, as dictated by the KZM. 
The presence of noise in the control field $g(t)$ gives rise to anti-KZ behavior, limiting the performance of adiabatic protocols. \label{fig:0}}
\end{figure}

The control over the system, however, is never perfect. In particular, the modulation in time of the control field might be subject to noise \cite{DKZ13,Ulm13,Pyka13,Lin14,Cui15}. In our study, we consider a thermally isolated system with no coupling to a thermal environment or heat bath, discussed in \cite{Patane08,Patane09,Rivas13,Viyuela14,Nalbach2015,Zhong,Braun,Yariv}. While the dynamics is described by unitary evolution, the external control field $g(t)$ includes stochastic fluctuations. 

Inspired by experimental protocols to test KZM \cite{DKZ13,Ulm13,Pyka13,Lin14,Cui15}, we consider the average over the realizations of noise and derive an exact master equation for the effectively open quantum many-body dynamics. 
As a result, the annealing dynamics is characterized by the interplay of two competing effects: (i) the approach to the adiabatic limit resulting from the universal suppression of diabatic excitations predicted by KZM and (ii) the accumulation of noise-induced excitations during the evolution. 
By a numerically exact approach, we show that for large values of the ramp time $\tau$, noise contributions dominate the dynamics and give rise to  anti-KZ behavior.  An important consequence for quantum annealing is the emergence of a finite optimal ramp time $\tau_{\rm opt}$ for the process.  The density of excitations can be reduced   by increasing the ramp time up to this optimal value  $\tau_{\rm opt}$, beyond which excitations accumulate, heating up the system. 
Explicitly, we argue that in the limit of small noise and finite time $\tau$, the noise-averaged density of excitations $n_W$ upon the completion of the annealing protocol is given by 
\begin{equation}\label{eq:n}
n_W\approx r\tau+c\tau^{-\beta},
\end{equation}
where $\beta$ is the universal KZM exponent, $c$ is a nonuniversal (dimensionful) prefactor, and $r$ is the rate at which the presence of noise in the control field generates excitations. 
The effective decoupling of the KZM dynamics from noise-induced effects  leads to the additive form of Eq. (\ref{eq:n}).
The (total) density of excitations  $n_W$ is minimized when ramp time is chosen to be
\begin{equation}
\label{tauopt}
\tau_{\rm opt}\propto r^{-1/(\beta+1)}.
\end{equation}
Furthermore, $r= \Lambda^2 W^2$, where $W^2$ characterizes the strength of the noise (in units of time) and $\Lambda$ sets the energy scale of the Hamiltonian.

We  verify our prediction by explicit calculations in the transverse-field Ising model. This exactly solvable model has been an important testbed for the quantum KZM \cite{ZDZ05,Dziarmaga05,Polkovnikov05,Dziarmaga10,Polkovnikov11,DZ13}. The Hamiltonian of the model is 
\begin{equation}
\label{HIsing}
H=-\sum_{n=1}^N\left(B\hat\sigma_n^x+J\hat\sigma_n^z\hat\sigma_{n+1}^z\right),
\end{equation}
where for simplicity we assume periodic boundary condition and an even number of spins $N$. This system  exhibits a quantum phase transition at $J=B$ between a paramagnetic phase ($B>J$) and a doubly degenerate ferromagnetic phase ($B<J$).
The model is of relevance to condensed matter systems \cite{E8} and can be realized in quantum simulators \cite{PC04,Islam11}.
Its quantum critical dynamics has recently been used in the laboratory to test the KZM in the quantum regime \cite{Cui15,Gong16}. 
The density of excitations generated by changing the ratio $B/J = t/\tau_Q$ in a time scale $\tau_Q$ matches the KZM prediction $n_0\sim 1/\sqrt{\tau_Q}$ corresponding to $d=\nu=z=1$, which leads to $\beta=1/2$ \cite{ZDZ05,Dziarmaga05,Polkovnikov05}.

We  consider the crossing of the QCP from the paramagnetic to the ferromagnetic phase of model~\eqref{HIsing}, following a quantum annealing scheme \cite{Farhi}. The time-dependent Hamiltonian is
\begin{equation}\label{eq:hamil}
H=-\Lambda\sum_{n=1}^N\left\{\left[1-g(t)\right]\hat\sigma_n^x+g(t)\hat\sigma_n^z\hat\sigma_{n+1}^z\right\},
\end{equation}
where $g=0$ ($g=1$) corresponds to the perfect paramagnet (ferromagnet) with a vanishing exchange coupling (magnetic field) and the QCP is at $g_c={1\over 2}$. Starting from the paramagnet
with $g(0)=0$, the control field $g(t)$ is varied to prepare the ground state  at $g(\tau)=1$. Hereafter, we set the energy scale $\Lambda$ to unity. In an adiabatic protocol, the paramagnetic phase is prepared free from excitations. 
 To account for the finite-time annealing dynamics generated by a noisy control field $g(t)$, we consider a prescheduled linear ramp 
 \begin{equation}\label{eq:g0}
g_0(t)=t/\tau,
\end{equation}
in the presence of a stochastic  perturbation $\gamma$,
\begin{equation}
\label{controlf}
g(t)=g_0(t)+\gamma(t), \quad 0<t<\tau,
\end{equation}
where $\gamma$ is a Gaussian white noise with zero mean and the second moment $\la\gamma(t)\gamma(t')\ra=W^2 \delta (t-t')$.
Here $W$ characterizes the strength of the noise (note that $\gamma$ is dimensionless and $W^2$ has units of time). White noise is a good approximation to ubiquitous colored noise with exponentially decaying correlations such as the Ornstein-Uhlenbeck process~\cite{Dalessio2013,Rahmani2013}. While our quantitative predictions may change for power-law correlated noise, we expect the slow decay of the noise spectrum (in the frequency domain) to allow for absorption of energy from the noisy drive and the aforementioned competition between adiabaticity and heating, qualitatively leading to the same anti-KZ behavior.

The linear ramp $g_0(t)$ of Eq.~\eqref{eq:g0} captures the physics of more general protocols that can be effectively linearized in the vicinity of the QCP, motivating this commonly used choice.
In particular, the annealing dynamics is directly related to the protocols considered in studies of the KZM, where only one coupling constant (e.g., $B/J$) is varied to traverse the critical point at $g_c=1/2$ in Eq.~\eqref{eq:hamil}. Neglecting the noise, the linear protocol~\eqref{eq:g0} crosses the QCP at $\delta t\equiv t-\tau/2=0$. Expanding in $\delta t$ gives $B(\delta t)/J(\delta t)=(\tau-2\delta t)/(\tau+2\delta t)=1-4\delta t/\tau+{\
\cal O}(\delta t^2/\tau^2)$.
As a result, for $\delta t/\tau\ll 1$ the ramp time $\tau$ is proportional to the quench time $\tau_Q$ \cite{Dziarmaga05} via $\tau_Q=\tau/4$. Thus, for $W=0$, we expect the dynamics above to generate an excitation density proportional to $1/\sqrt{\tau}$.

In what follows, we focus on the dynamics generated by Eqs. (\ref{eq:hamil}), \eqref{eq:g0}, and (\ref{controlf}). First we make use of Novikov's theorem \cite{Novikov65} to show that the  noise-average density matrix obeys an effectively open quantum dynamics described by a Lindblad-type master equation.
To this end we separate the deterministic and stochastic parts of the time-dependent noisy Hamiltonian
\begin{equation}
H(t)=H_0(t)+\gamma(t) V,
\end{equation}
where from Eq.~\eqref{eq:hamil} we find
\beqa
H_0(t)&=&-\sum_{n=1}^N\left\{\left[1-g_0(t)\right]\hat\sigma_n^x+g_0(t)\hat\sigma_n^z\hat\sigma_{n+1}^z\right\},\\
V&=&-\sum_{n=1}^N\left (-\hat\sigma_n^x+\hat\sigma_n^z\hat\sigma_{n+1}^z\right ).
\label{Hsplit}
\eeqa
For each realization of noise $\gamma(t)$, assuming we start from a pure state $|\psi(0)\rangle$, the time evolution is governed by a stochastic Schr\"odinger equation,  
\begin{equation}
\label{eq:schr}
i{d\over dt}|\psi(t)\rangle=\left[H_0(t)+\gamma(t) V\right]|\psi(t)\rangle.
\end{equation}
It is possible for the system to decohere at long times of evolution, violating the assumption of unitarity for each realization of noise. As we are interested in the limit of small noise in systems with high thermal isolation, the fate of the system for $t\geqslant \tau_{\rm opt}$ does not affect our prediction of the anti-KZ behavior and the scaling~\eqref{tauopt} for $\tau_{\rm opt}$.

As shown in Ref. \cite{SM} using Novikov's theorem~\cite{Novikov65} (see also Refs.~\cite{Pilcher2013,Rahmani2015}), Eq. (\ref{eq:schr}) unravels the following nonperturbative exact master equation for the  noise-averaged density matrix $\rho(t)$, 
\begin{eqnarray}
{d\over dt}\rho(t)&=&-i[H_0(t),\rho(t)]-{{W^2}\over 2}[V,[V,\rho(t)]].
\label{eq:Novikov}
\end{eqnarray}
The first term in the right-hand side accounts for the unitary evolution generated by the prescheduled Hamiltonian $H_0(t)$.
The second one induces an effectively open quantum dynamics \cite{BP02} with Hermitian bath operators $V$ that include both one- and two-body spin interactions; see Eq. (\ref{Hsplit}).

We next use a Jordan-Wigner transformation to write the stochastic Hamiltonian in terms of free fermions as~\cite{Dziarmaga05}
\begin{equation}
\label{eq:Hn}
H=-\sum_{n=1}^N\left\{g(t)c^\dagger_nc_{n+1}+g(t)c_{n+1}c_n-\left[1-g(t)\right]c^\dagger_nc_n+{\rm H.c}\right\},\nonumber\\
\end{equation}
where $c_n$ are fermionic annihilation operators and we have dropped an irrelevant time-dependent real function. We note that (\ref{eq:Hn}) commutes with the fermion parity operator for each noise  realization, i.e., $[H(t),P]=0$, where $P=\prod_n(1-2c^\dagger_nc_n)$.  We work in the even parity subspace where  boundary condition are antiperiodic, i.e., $c_{N+1}=-c_1$. Using a Fourier transformation $c_n=e^{-i\pi/4}\sum_k c_k e^{ikn}/\sqrt{N}$ with
$k=\pm (2m-1){\pi\over N}$ for  $m=1,2,\dots, N/2$,
we find
\begin{equation}
\label{eq:Hk_ann}
H(t)=2\sum_{k>0}\Psi_k^\dagger\left\{\sigma_z\left[1-g(t)-g(t)\cos k\right]+\sigma_xg(t)\sin k\right\}\Psi_k,
\end{equation}
where $\Psi^\dagger_k\equiv \left(c^\dagger_k,c_{-k}\right)$. The Pauli matrices used in writing the above Hamiltonian in momentum space are not to be confused with the original spin operators indicated by $\hat \sigma^{x,y,z}_n$. Both $H_0(t)$ and $V$ can then be written in the form
\begin{eqnarray}
H_0(t)&=&\sum_{k>0}\Psi_k^\dagger\left[\sigma_zh_k^z(t)+\sigma_xh_k^x(t)\right]\Psi_k,\\ V&=&\sum_{k>0}\Psi_k^\dagger\left[\sigma_zv_k^z+\sigma_xv_k^x\right]\Psi_k,
\end{eqnarray}
 with
$h_k^z(t)=2\left[1-g_0(t)-g_0(t)\cos k\right]$, $h_k^x(t)=2g_0(t)\sin k$, $v_k^z=-2\left(1+\cos k\right)$, $v_k^x=2\sin k$. 
The initial state is taken to be the ground state of the Hamiltonian~\eqref{eq:Hk_ann} for $g=0$, which is simply given by $|\psi(0)\rangle=\bigotimes_{k>0}|0\rangle_k$, where $|0\rangle_k$ is the vacuum in one sector of the block-diagonal Hamiltonian (each positive momentum $k$ is only coupled to $-k$).
The noise-averaged density matrix ${\rho(t)}$ can be computed by solving equation of motion~\eqref{eq:Novikov} with the initial condition $\rho(0)=|\psi(0)\rangle\langle\psi(0)|$. {We neglect the terms in the master equation that couple different modes so the solution retains a tensor-product form  $\rho(t)=\bigotimes_{k>0}\varrho_k(t)$. As argued in Ref.~\cite{Nalbach2015}, this approximate density matrix yields exact noise-averaged expectation values for quadratic operators with translation symmetry such as Eqs.~\eqref{eq:n_W} and \eqref{eq:Q} \cite{note}.
 Our results for the quartic operator~\eqref{eq:DE}, however, are approximate.}
Fermion parity is conserved in each sector and the corresponding initial state $|0\rangle_k$ has even parity. Thus, the single-mode Hilbert space is spanned by the pseudospins $|\uparrow\rangle_k=c_k^{\dagger}c_{-k}^{\dagger}|0\rangle_k$ and  $|\downarrow\rangle_k=|0\rangle_k$ (for any $2\times 2$ matrix $O$, the quadratic operator $\Psi^\dagger_k O\Psi_k$ in the above basis is given by the matrix $O$). The $k$-mode density matrix $\varrho_k(t)$ can then be represented as a two-dimensional matrix in the above basis, whose evolution is dictated by the master equation
\begin{eqnarray}\label{eq:matrix_master}
{d\over dt}\varrho_k(t)&=&-i[\sigma_z h_k^z(t)+\sigma_x h_k^x(t),\varrho_k(t)]\nonumber\\
&&-{{W^2}\over 2}[\sigma_z v_k^z+\sigma_x v_k^x,[\sigma_z v_k^z+\sigma_x v_k^x,\varrho_k(t)]].
\end{eqnarray}
The dissipative part induces dephasing in each mode along the $x$ and $z$ directions and contains as well mixed $xz$ and $zx$ terms.
 \begin{figure}[t]
\centering
		\includegraphics[width=\columnwidth]{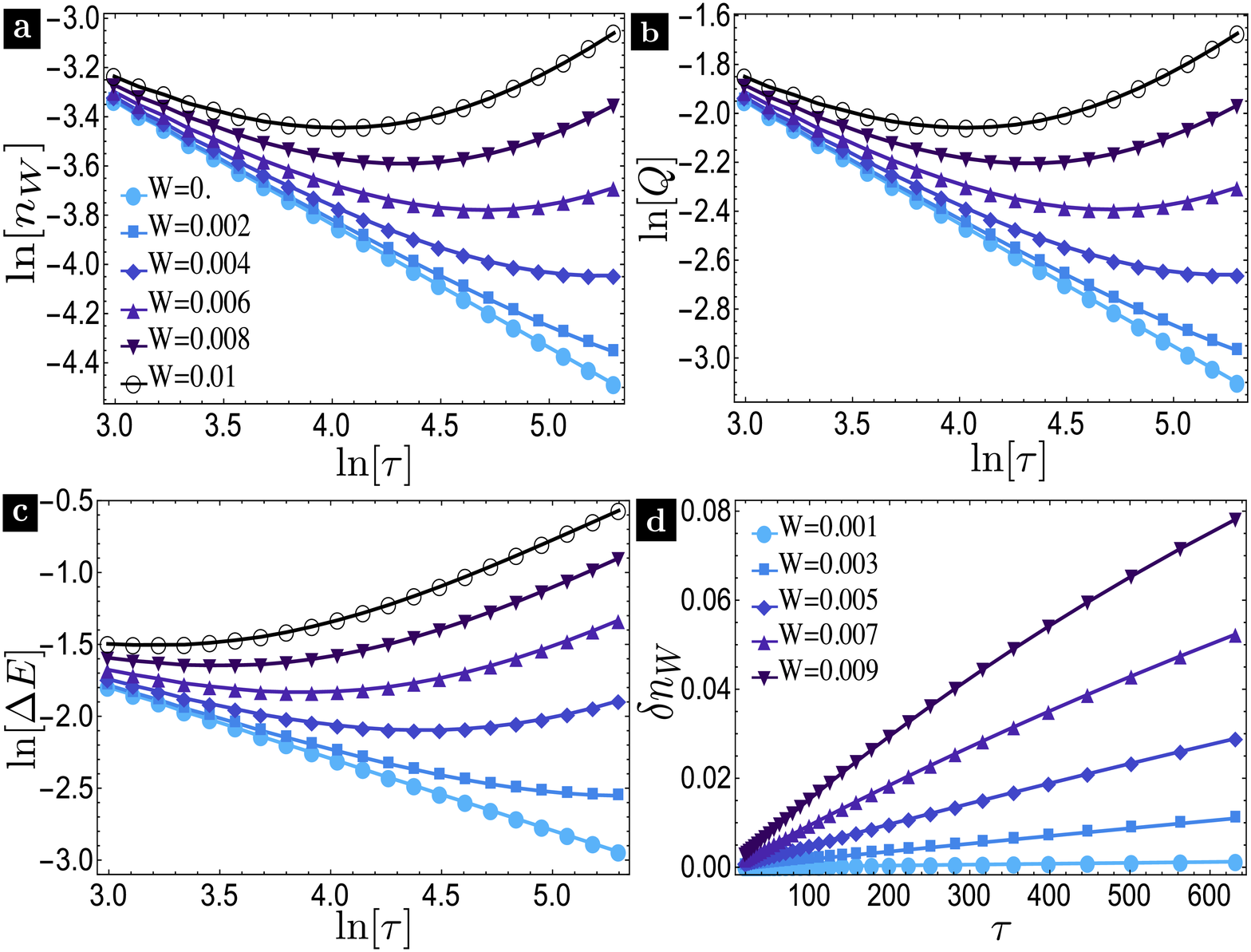}
	\caption{{\bf Anti-Kibble-Zurek behavior induced by a noisy control field.}  {\bf (a)} The density of excitations upon completion of the annealing schedule for different values of $W$ as a function of ramp time $\tau$. The numerically exact value surpasses the power-law scaling predicted by the KZM when $W^2>0$. {\bf (b)} A similar dependence is observed in the residual energy density $Q$. {\bf (c)} Noise-induced effects are more pronounced in the energy spread $\Delta E$ and already manifested  for short ramp times. {\bf (d)} The difference between the density of excitations generated in the presence and absence of noise  scales linearly as a function of ramp time $\tau$ for fast ramps with $W^2\tau <1$, at a characteristic heating rate.  Deviations are noticeable for long ramps.
	\label{fig:2}}
\end{figure}
We resort to the numerical solution of this set of master equations and characterize the nonadiabatic open dynamics in terms of three different observables.
The density of excitations is given by
\beqa\label{eq:n_W}
n_W=1-{1\over N}\sum_{k>0} \la G_k(\tau)|\varrho_k(t)|G_k(\tau)\rangle,
\eeqa
where $|G_k(\tau)\rangle$ is the $k$-sector ground state.
Further, we compute the mean excess energy of the final state over the corresponding ground state:
\beqa\label{eq:Q}
Q=\left\{\tr[\rho(\tau)H(\tau)]-\langle G(\tau)|H(\tau)|G(\tau)\rangle\right\}/N.
\eeqa
This so-called residual energy density is known to scale universally for $W=0$ as a function of the ramp time in a KZM quench \cite{DeGrandi10}.
Finally, the energy spread  of the final state~\cite{Bunin2011,Dalessio2013}
\beqa\Delta\label{eq:DE} E=\left\{\tr[\rho(\tau)H^2(\tau)]-\tr[\rho(\tau)H(\tau)]^2\right\}^{1/2}/N
\eeqa
 per particle is obtained in a similar manner.  For numerical simulations we choose  the system size  $N=1024$ for which  there are no important finite-size effects and a range of values for the ramp time where universal KZM scaling is observed for $W=0$.

 The dependence of the density of excitations $n_W$ on the ramp time  $\tau$ is shown in   Fig.~\ref{fig:2}(a) for several values of the noise strength $W$. For short ramp times $\tau$, the effect of noise in the control field is negligible and the density of excitations  scales as a power law in agreement with the KZM prediction, $n_W\propto\tau^{-1/2}$. For longer ramp times, noise-induced effects  dominate the nonadiabatic dynamics, leading to the growth of $n_W$ with the ramp time $\tau$ (for fixed $W$). This is  the anti-KZ regime, where decreasing the annealing rate $1/\tau$  results in a higher excitation of the system. 
 \begin{figure}
\centering
\includegraphics[width=0.7\columnwidth]{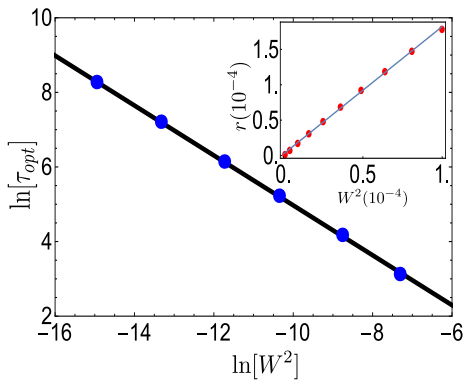}
\caption{{\bf Universal scaling of the optimal annealing time. }
 The scaling of the optimal annealing time  $\tau_{\rm opt}$ that minimizes the density of excitations as a function of heating rate $r$, verifies the universal prediction in Eq. (\ref{tauopt}). {\bf Inset:} The heating rate $r$ is shown to scale linearly as a function of square of the strength of noise $W^2$.
\label{fig:3}
}	
\end{figure}
 In the limit of very long times, $n_W$ is completely governed by the anti-KZ contribution and approaches $1/2$, as confirmed by the  analytical estimate derived  in Ref. \cite{SM}.

The interplay between the approach to adiabaticity and the accumulation of noise-induced excitations also appears in other observables, including the residual mean energy $Q$. For $W=0$, $Q$ scales as the density of defects in our model and for $W^2>0$ an analogous anti-KZ regime is observed after surpassing an optimal ramp time [see Fig \ref{fig:2}(b)].
The optimal time does depend on the observable. In particular, energy fluctuations exhibit the anti-KZ behavior at earlier stages of the dynamics, as shown in Fig \ref{fig:2}(c). We can define the difference between the number of excitations generated in the presence and absence of noise (with strength $W$),  $\delta n_W=n_W-n_0$. 
The data in Fig \ref{fig:2}(d) confirm that for moderate ramp times and small values of $W$m noise-induced generation of excitations is well characterized by a heating rate $r$, supporting Eq.~\eqref{eq:n}.  In this regime, the excess of excitations exhibits a linear growth  $\delta n_W \simeq r \tau$. Deviations from it as a function of the ramp time are expected and observed for longer ramps.  

We next demonstrate the validity of the scaling ansatz (\ref{eq:n}) relating the optimal ramp time to the noise strength. 
The heating rate is set by the amplitude of the noise fluctuations $W^2$ as seen in Fig.~\ref{fig:3}. Ultimately, this scaling is dictated by the quantum speed limits for open systems \cite{Taddei13,delcampo13}. Minimizing the density of excitations as a function of the ramp time for different noise strengths,  we determine the scaling of the optimal ramp time  on $W$. A linear fit to the data shows that  the density  of excitations is minimized when the ramp time is $\tau_{\rm opt}=a (W^2)^b$ with $a=0.193\pm0.003$ and $b=-0.669\pm 0.004$ in excellent agreement with the theoretical prediction
 $\tau_{\rm opt}\propto (W^2)^{-2/3}$, e.g., following Eq. (\ref{tauopt}).
This sets an upper limit to the ramp time in quantum annealing protocols, above which  anti-KZ behavior dominates, and the density of excitations increases with the ramp time.

In conclusion, we have provided a natural mechanism to explain the anti-Kibble-Zurek behavior in the quantum critical dynamics of a thermally isolated system driven by a noisy control fields. Our results show the limits to adiabatic strategies in quantum annealing and indicate that the optical annealing time follows a universal scaling law as a function of the amplitude of the noise fluctuations.

It is a pleasure to thank N. A. Sinitsyn for illuminating discussions. This work was supported by UMass Boston (project P20150000029279) (A.D. and A.dC.), NSERC (A.R.), and Max Planck-UBC Centre for Quantum Materials (A.R.).


\begin{thebibliography}{99}
\bibitem{note} We have checked this explicitly by solving the many-body master equation for small systems and by the numerically-exact integration of the equations of motion for the expectation value of observables in the Heisenberg picture.
\end{thebibliography}
\end{document}


\title{
 Supplemental Material for ``Anti-Kibble-Zurek Behavior in Crossing the Quantum Critical Point of a Thermally Isolated System Driven by a Noisy Control Field''
}

\author{Anirban Dutta}
\affiliation{Department of Physics, University of Massachusetts, Boston, MA 02125, USA}
\author{Armin Rahmani}
\affiliation{Department of Physics and Astronomy and Quantum Matter Institute, University of British Columbia, Vancouver, British Columbia,  Canada V6T 1Z4}
\author{Adolfo del Campo}
\affiliation{Department of Physics, University of Massachusetts, Boston, MA 02125, USA}

\maketitle
\newcommand{\be}{\begin{equation}}
\newcommand{\ee}{\end{equation}}
\newcommand{\bea}{\begin{eqnarray}}
\newcommand{\eea}{\end{eqnarray}}

\def\q{{\bf q}}

\def\G{\Gamma}
\def\L{\Lambda}
\def\la{\lambda}
\def\g{\gamma}
\def\al{\alpha}
\def\s{\sigma}
\def\e{\epsilon}
\def\k{\kappa}
\def\ve{\varepsilon}
\def\l{\left}
\def\r{\right}
\def\te{\mbox{e}}
\def\d{{\rm d}}
\def\t{{\rm t}}
\def\K{{\rm K}}
\def\N{{\rm N}}
\def\H{{\rm H}}
\def\la{\langle}
\def\ra{\rangle}
\def\om{\omega}
\def\Om{\Omega}
\def\vep{\varepsilon}
\def\wh{\widehat}
\def\tr{{\rm Tr}}
\def\da{\dagger}
\def\iz{\left}
\def\zi{\right}
\newcommand{\beq}{\begin{equation}}
\newcommand{\eeq}{\end{equation}}
\newcommand{\beqa}{\begin{eqnarray}}
\newcommand{\eeqa}{\end{eqnarray}}
\newcommand{\intf}{\int_{-\infty}^\infty}
\newcommand{\into}{\int_0^\infty}

\section{Derivation of the master equation for the noise-averaged density matrix}
Stochastic Hamiltonians often appear in quantum optics and quantum measurement theory and  it is in this context that effective master equations for the noise-average density matrix have been discussed in \cite{Milburn91,Moya93}. 
A rigorous derivation relies on the use of Novikov's theorem to evaluate the noise average of a functional of a stochastic variable \cite{Novikov65,Budini01} (see Appendix. A of Ref.~\cite{Rahmani2015} for an alternative approach).
Generally, it is only possible to do so  in a perturbative treatment, up to correction of order $\mathcal{O}(W^4)$ \cite{Budini01}.
We focus on the dynamics of the prescheduled deterministic Hamiltonian  $H_0$ in the presence of an stochastic perturbation $ \gamma V$ with a single real Gaussian white noise $\gamma$ that represents fluctuations in the control field.
In this scenario our derivation is  nonperturbative in the strength of the noise $\gamma$ and the master equation is exact. Using the stochastic Sch\"rodinger equation
\begin{equation}
i{d\over dt}|\psi(t)\rangle=\left[H_0(t)+\gamma V\right]|\psi(t)\rangle,
\end{equation}
and the definition of the stochastic density matrix  $\rho_{\gamma}(t)=|\psi_{\gamma}(t)\rangle\langle\psi_{\gamma}(t)|$, the corresponding Heisenberg equation reads
\begin{equation}
{{d}\over{dt}}\rho_{\gamma}=-i[H_0(t),\rho_{\gamma}]-i[V,\gamma \rho_{\gamma}].
\end{equation}
We next perform an average over noise realizations and denote the noise-averaged density matrix by $\rho=\la  \rho_{\gamma}\ra$, that is a solution of the master equation
\begin{equation}
{{d}\over{dt}}\rho=-i[H_0(t),\rho]-i[V,\langle {\gamma} \rho_{\gamma}\rangle].
\end{equation}
The second term in the right-hand side can be evaluated 
using Novikov's theorem  \cite{Novikov65} 
\begin{equation}
\langle\gamma(t) F[\gamma]\rangle={1\over2}\left\langle{{\delta F}\over{\delta \gamma(s)}}\right\rangle_{s=t}
\end{equation}
to find
\begin{equation}
\langle \gamma \rho_\gamma\rangle=-{{iW^2}\over{2}}[V,\rho].
\end{equation}
As a result, the master equation takes the form 
\begin{eqnarray}
{d\over dt}\rho(t)&=&-i[H_0(t),\rho(t)]+\mathcal{D}[\rho(t)],
\label{ME}
\end{eqnarray}
where the dissipator is given by
\beqa
\label{Diss}
\mathcal{D}[\rho(t)]=-{{W^2}\over 2}[V,[V,\rho(t)]].
\eeqa
Further, this master equation can be rewritten in the standard Lindblad form \cite{Lindblad76,BP02}
\beqa
\mathcal{D}[\rho]=W^2\left(V\rho V^\dag-\frac{1}{2}\rho V^\dag V- \frac{1}{2} V^\dag V\rho\right),
\eeqa
with Hermitian bath operators $V=V^\dag$, and generates a  Markovian dynamics as $W^2>0$.
Equations (\ref{ME})-(\ref{Diss}) determine the master equation used in the manuscript.

\section{Asymptotics of anti-Kibble-Zurek scaling for long ramp times }

The availability of analytical results for modified Landau-Zener ramps under nonunitary dynamics is severely restricted.
Prominent results include the Kayanuma formula for fast diagonal noise \cite{Kayanuma} as well as the Pokrovsky-Sinitsyn scaling for more general cases \cite{PS03,PS04}.
In this appendix we aim at providing an analytical estimate of the noise-averaged density of excitations generated during the quantum critical dynamics of the transverse-field Ising model. In particular, we wish to provide analytical evidences of the additive nature of the KZM and noise-induced contributions to the density of excitations and the saturation of the latter at long times of evolution.  The form of the $k$-mode master equation precludes us from finding exact results and we next adopt a simplified scheme that arises when the Ising chain is driven by a linear ramp of the magnetic field while the strength of the interactions is kept constant. Using the mode decoupling approximation, the dynamics of each $k$-mode is generated by a two-level Hamiltonian of the form
\begin{eqnarray}
H=g(t)\sigma^{z}+\Delta\sigma^{x}.
\end{eqnarray}
In the absence of noise, for $W=0$,  $g(t)$ varies linearly with time $g(t)=v t$ and the dynamics corresponds to a Landau-Zener crossing. 
Consider the system is prepared at $t=-\infty$ in the corresponding ground state of the above Hamiltonian, which is the positive energy eigenstate of the Pauli matrix $\sigma^z$. 
From the exact time evolution of the initial state, it is possible to evaluate the excitation probability after the avoided crossing  as $t\rightarrow\infty$ via the Landau-Zener formula 
is given by 
\begin{eqnarray}
P_{LZ}=e^{-2\pi \Delta^2/|v|}.
\label{eq:LZ}
\end{eqnarray} 	
To account for the dynamics of the transverse-field Ising model, we set $\hbar=1$, lattice constant $|a|=1$ and Ising coupling $J=1$. 
The excitation probability $P_{LZ}^{(k)}$ in  each $k$-mode of the Ising chain can be estimated with the identification \cite{Dziarmaga05,Polkovnikov05}
\beqa
\Delta^2/|v|=\tau k^2.
\eeqa
\begin{figure}[t]
	\begin{center}
		\includegraphics[width=0.49\columnwidth]{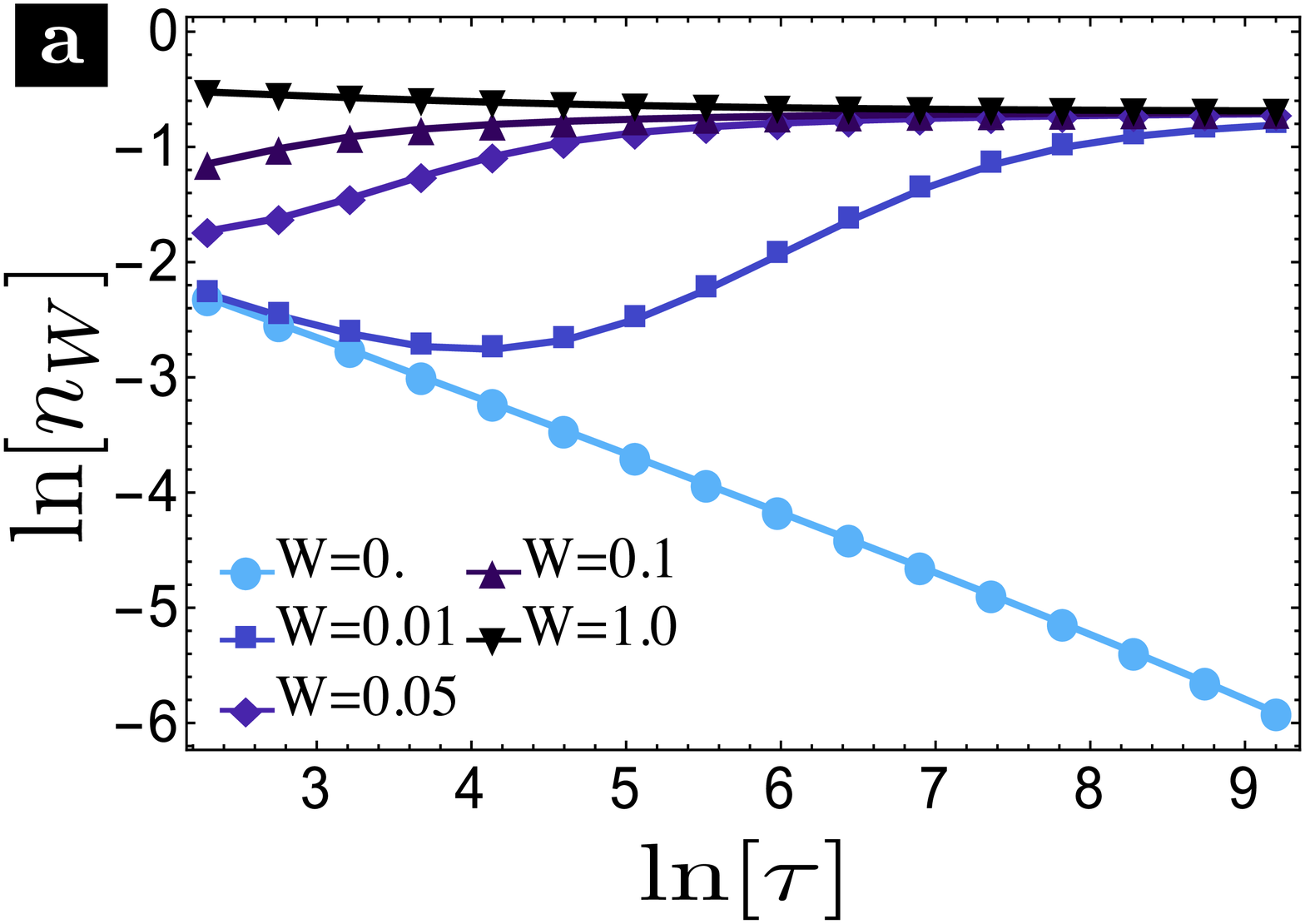}
		\includegraphics[width=0.49\columnwidth]{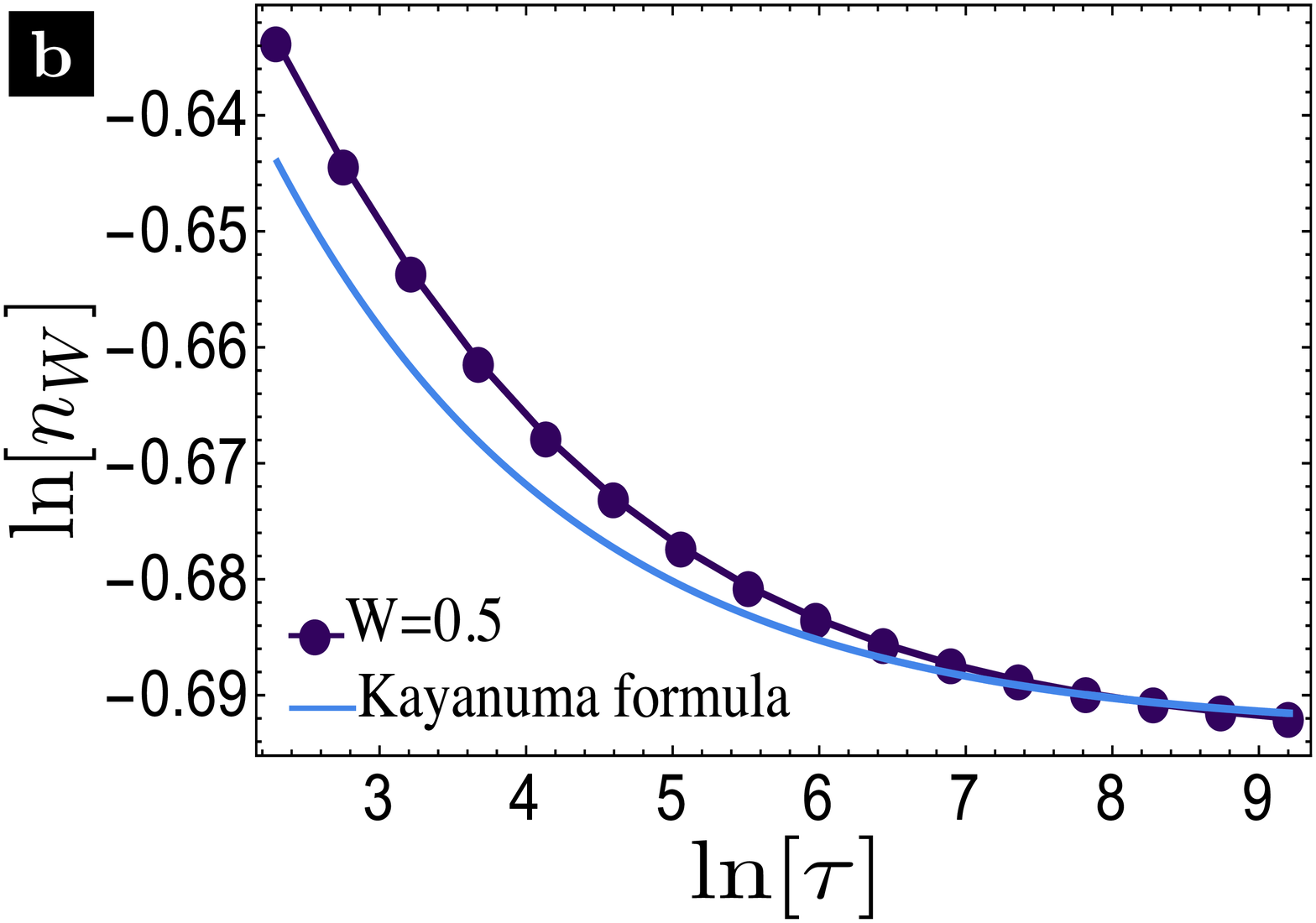}
	\end{center}
	\caption{{\bf Asymptotic anti-Kibble-Zurek dynamics.} {\bf (a)} The final density of defect for different values of $W$ as a function of total ramp time $\tau$. The date is for a linear ramp from $g=2$ to $g=0$ in the limit of very slow quenches where the density of excitations is governed by the anti-KZ behavior and approaches $1/2$. {\bf (b)} Comparison of the density of excitations obtained from the numerically-exact solution and the analytical scaling derived via the Kayanuma formula (continuous line without dots) Eq.~\eqref{eq:kaya_formula}.
		\label{fig5}}
\end{figure}
In presence of fast diagonal noise, the Landau-Zener prediction is to be modified, and it is replaced by the  Kayanuma formula \cite{Kayanuma}, according to which the  excitation probability reads
\begin{eqnarray}
P_{K}={{1}\over{2}}(1+e^{-4\pi \Delta^2/|v|}).
\label{eq:kayanuma1}
\end{eqnarray}
Integrating over the Brillouin zone  and taking the continuum limit, the density of excitations can be estimated,
\begin{eqnarray}
n_{K}&=&\frac{1}{N}\sum_{k> 0} P_K^{(k)}
\simeq 
{{1}\over{2}}+{{1}\over{4 \pi}}{{1}\over{\sqrt{\tau}}}.
\label{eq:kaya_formula}
\end{eqnarray} 

Figure \ref{fig5}(a) shows the long-time dynamics of the density of excitations in the transverse-field Ising model, where the noise-induced anti-KZ behavior saturates at $1/2$.
The fixed point of the dissipator $\mathcal{D}[\cdot]$ that describes the asymptotic density matrix in the absence of unitary driving is given by the completely mixed (infinite-temperature) state 
\beqa
\rho(t)\sim\bigotimes_k\textbf{I}_2/2,
\eeqa 
where $\textbf{I}_2$ is the identity matrix in the eigenbasis of the final $k$-mode Hamiltonian.
For long but finite ramp times, excitations build up over this asymptotic value  following a universal KZ scaling, according to (\ref{eq:kaya_formula}), see 
Fig. \ref{fig5}(b).
%
%